# Effect of thickness on the maximum potential drop of current collectors

Jose Miguel Campillo-Robles *, Xabier Artetxe and Karmele del Teso Sánchez

*Mekanika eta Ekoizpen Industrialeko Saila, Mondragon Unibertsitatea, Loramendi 4, 20500 Arrasate, Basque Country, Spain*
**\*** Corresponding author. Tel.: +34 943 797 400

E-mail address: jmcampillo@mondragon.edu

The basic principle for achieving high-power capability on an electrochemical energy storage cell is minimizing the overall resistance. The resistance due to current collecting systems has not received sufficient attention in the past, presumably because it was not considered of significance for low-power batteries and supercapacitors. However, the necessity of high-power cells has reduced the other sources of inner resistance, and current collector potential drop has become more important. Moreover, the miniaturization of energy storage devices could increase the ohmic loses in current collectors. In this work, we have developed an electrical model to assess the effect of current collector thickness on the maximum potential drop. We have found that the thickness of current collectors is a critical parameter that can increase the maximum potential drop drastically. Indeed, maximum potential drop of current collectors remains almost constant for thicknesses greater than 500 μm, but below this value, there is an inverse relationship between the maximum potential drop and the thickness. We have also analyzed the effect of material and tab position in the maximum potential drop.

Recent successes in the effort to miniaturize electronic devices have driven the need for very thin batteries and supercapacitors. As a result, there is an increasing tendency to the miniaturization and integration of energy and power sources for wearable and portable devices[1-3]. Indeed, thin film batteries and supercapacitors have important applications in a variety of consumer and medical products. This kind of energy and power sources are assuming new form factors, becoming ultra-thin, flexible, stretchable, etc. For this reason, they require new features, designs and manufacturing[4,5], which traditional technologies are unable to provide. However, it is well known that the size of an object can have direct effect on its properties. Therefore, these devices have received increasing attention from both academia and industry during the last decade.

Current collecting systems have three important functions in energy storage devices. First, they are the physical support structure to hold the active materials in place inside the cell (otherwise, it would be a very brittle structure). Second, current collectors absorb mechanical stresses caused by external forces or volume changes of active materials during cycling. Finally, the most important function, they carry the electric current in or out the electrodes. Researchers have devoted time to develop better current collecting systems, most of the times using numerical analysis. For example, several improved grids have been proposed for the widely used lead acid batteries[6-11]. Extended research has been performed also to understand the inner conduction phenomena in Lithium Ion Batteries (LIBs)[12]. Typically, a LIB cell contains a positive current collector, an aluminum (Al) sheet, and a negative current collector, a copper (Cu) sheet, with active material layers coated on them. They account for about 15-20 % by weight and 10-15% by cost of a LIB[13]. For the industrial and consumer point of view, thinner, lighter and cheaper current collectors would be preferred. Last years, researchers have performed numerical analysis to investigate the electrical behavior of inner elements of LIBs[14-18]. In these works, electrical behavior of current collectors with different geometries has been examined. Taheri *et al*. performed a very exhaustive electrical analysis of different design parameters related to current collectors of large-scale LIBs[14]. However, the effect of current collector thickness on the electrical behavior has not been evaluated yet. Thus, the main goal of this research is to study one very practical problem of LIBs or supercapacitors: the effect of thickness in the electrical behavior of current collectors under usual working conditions.

We have developed a deliberately simple electrical model to calculate the electrical potential, $V$, generated by a crossing current density, $\vec{j}$, in four metallic thin sheets. The electrical potential can be obtained solving the following two equations:

$$\vec{j} = -\left(\frac{1}{\rho}\right)\vec{\nabla}V, \qquad (1)$$

$$\vec{E} = -\vec{\nabla}V, \qquad (2)$$

where $\rho$ is the electrical resistivity of the metallic material and $\vec{E}$ the electrical field. We have supposed a divergence free current density and no charge accumulation in the system. The governing differential equations for the charge balance have been solved numerically using COMSOL Multiphysics® (version 5.3) with the AC/DC toolbox[19]. We have used a physics controlled tetrahedral adaptive mesh in all the calculations. In addition, meshes with different densities were used in order to ensure the accuracy and the mesh independency of the solutions.

The dependent variable, electric potential, has been discretized using second order elements. The maximum number of tetrahedral elements used in the simulations has been $2.5 \cdot 10^6$. Moreover, a stationary linear solver was chosen, with a tolerance for convergence below $10^{-3}$ for all variables. Finally, a Dirichlet boundary condition for the electric potential has been applied on the intensity outlet surface of all the simulated systems, top of the metallic collectors, $V = 0$.

To perform the numerical analysis, we have taken as a reference the current collectors of a commercial Lithium Ion Capacitor (LIC) pouch cell: 1.100 F, JM Energy[20]. In this pouch cell, there are 17 double-sided electrode-pairs, with current collectors of Cu for anode (15 μm thickness) and Al for cathode (30 μm thickness). These current collectors can be divided in two different parts: a rectangular sheet, $b = 10$ cm and $L = 12.8$ cm, and a collecting tab, $a = 5$ cm and $c = 1$ cm. We have used these lengths to build the four geometries of metallic sheets simulated in this work. We have started simulating two rectangular sheets of ($b \times L$) dimensions [see Figs. 1 (a) and (b)]. Nevertheless, to construct real current collectors, we need to add to the rectangular sheet a collecting tab of ($a \times c$) dimensions. Figs. 1 (c) and (d) show current collectors with the collecting tab positioned in two different positions in the upper part of the sheet. The current collector of the commercial LIC pouch cell is that shown in Fig. 1 (c). The thickness, $d$, of these four metallic sheets has not been plotted, but it is perpendicular to the figures. We have performed the numerical analysis using different values for the thickness: 10 μm, 50 μm, 100 μm, 500 μm, 1 mm and 5 mm. The thickness of commercial current collectors is usually among these values.

We have used common intensities of pouch LIC cells to fix the working conditions of the simulated current collectors: 50 A, 100 A, 200 A and 360 A. Therefore, the intensity $I$ flowing through each current collector can be calculated dividing the previous intensities by 17. As a result, the corresponding $I$ intensities are the next ones: 2.94 A, 5.88 A, 11.76 A and 21.17 A. For all these intensities, we have calculated the maximum potential drop inside the four metallic sheets of Fig. 1. The inlet electric current distribution has been modelled in two different ways. In the current collector of Fig. 1 (a) a uniform current density of value $I/(bd)$ crosses from the bottom section ($b \times d$) to the upper part of the rectangular sheet. In the other three cases [see Figs. 1 (b), (c) and (d)], the uniform current intensity of value $I/(2bL)$ enters into the current collectors by the two symmetric side faces ($b \times L$) of the rectangular sheet. This last distribution of the electric current is the usual one in current collectors.

Several studies have shown that reducing one of the dimensions of a conductor will change the electrical resistivity of the material[21]. Metallic thin films of thickness lower than 100 nm show an increasing resistivity with the reduction of thickness, due to electron scattering phenomena[22,23]. Nevertheless, the exact value of the critical thickness, in which the resistivity starts increasing from the bulk value, depends on the chemical element of the film[24]. Anyway, the thickness of commonly used battery and supercapacitor current collectors is greater than 100 nm; they are usually on a scale of micrometers. Therefore, a constant bulk resistivity can be assumed for them. In this work, we have taken $16.7 \times 10^{-9}$ Ω m for the electrical resistivity of Cu and $26.5 \times 10^{-9}$ Ω m for Al, both at 20 °C[25].

First, we have analyzed a rectangular Cu sheet of $L$ height, $b$ width and $d$ thickness [see Fig. 1 (a)]. As explained before, in this metallic sheet the current density $I/(bd)$ flows upwards from the bottom section ($b \times d$). Therefore, the current density has the same value at all heights.

According to the Ohm's law, we can obtain an analytical expression for the maximum potential drop, $V_{max}$, which is expressed directly as follows:

$$V_{max} = \frac{\rho L I}{bd} \propto \frac{1}{d}. \qquad (3)$$

From this relation, we can obtain the resistance of the rectangular sheet for Cu and Al as a function of the thickness: $R_{Cu}(m\Omega) = 21.34/d(\mu m)$ and $R_{Al}(m\Omega) = 33.92/d(\mu m)$. Al sheet shows a greater resistance than Cu sheet, due to the higher resistivity of Al, generating more heat by means of Joule effect.

Figure 1 (a) also shows the potential distribution and the current streamlines obtained from the electrical simulations of the rectangular sheet. As it can be expected from Eq. (3), the electric potential changes linearly through the height of the sheet, with a maximum placed in the bottom of the sheet. Furthermore, the electric current streamlines are straight lines. Equipotential lines are not shown in this figure, but they are horizontal lines, perpendicular to streamlines. We have also plotted $V_{max}$ as a function of the thickness for different intensities [see Fig. 2 (a)]. All the trends follow the same relationship described in Eq. (3). The maximum potential drop of a rectangular sheet with a selected thickness shows a linear increment with the intensity. For thicknesses greater than 500 µm, $V_{max}$ remains almost constant, corresponding to the bulk value. For thicknesses lower than 500 µm, $V_{max}$ increases with the reduction of the thickness of the current collector following Eq. (3). Moreover, a logarithmic relation between the maximum potential drop and the thickness of the rectangular sheet can be derived directly from Eq. (3):

$$\ln(V_{max}) = -\ln(d) + C, \qquad (4)$$

where $C$ is a constant dependent on the system parameters. Figure 3 (a) shows a linear fit of the logarithmic relation ($R^2 = 1$).

We have analyzed a second case for the same rectangular sheet of Cu. In this second case, we have only changed the inlet surface of the current density. As mentioned before, the uniform current density $I/(2bL)$ enters into the sheet by the two symmetric side faces of the rectangular sheet ($b \times L$) [see Fig. 1 (b)]. As in the first case, we can obtain an analytical expression for $V_{max}$:

$$V_{max} = \frac{\rho L I}{2bd} \propto \frac{1}{d}. \qquad (5)$$

$V_{max}$ is half the value obtained in the previous case [Eq. (3)], and it remains showing an inversely proportional relationship to the thickness of the sheet. Even though, $V_{max}$ and thickness show the same relation, the potential gradient is not uniform all across the height of the rectangular sheet, as it happens in the first case [see Figs. 1 (a) and (b)]. The current density inside the sheet increases with the height of the sheet, and as a result, the potential gradient increases with the height too. In this second case, the current streamlines and equipotential lines (not shown) are also vertical and horizontal straight lines.

These previous geometries are not appropriate to collect the current in commercial LIBs and supercapacitors. It is necessary to add a collector tab to connect all the electrodes of the same polarity. For this reason, we are going to simulate the geometry of the current collectors of 1.100 F JM Energy pouch cell [see Fig. 1 (c)]. In this case, the collecting tab is in the middle of the rectangular sheet (symmetric configuration). We have supposed a uniform inlet current density $I/(2bL)$ entering the current collector by the two lateral sides of the rectangular sheet. In Fig. 1 (c), potential distribution, equipotential lines and current streamlines obtained from the numerical analysis have been plotted. The equipotential lines and current streamlines are not straight lines, but they show a symmetric pattern around the symmetry vertical plane of the current collector. Furthermore, the potential gradient is greater near the collecting tab. As mentioned in the second case of the rectangular sheet, in-plane current density increases with height and this produces a greater gradient in the upper part of the current collector. Moreover, the convergence of the streamlines at the collecting tab creates the so-called constriction resistance, which also increases the potential gradient. Comparing the maximum potential drop with the two previous cases of rectangular sheets (same thickness and intensity), the current collector geometry shows lower $V_{max}$ than the first case of the rectangular sheet [see Fig. 2], but it is 50 % greater than the second case of the rectangular sheet. Therefore, the addition of the collecting tab duplicates the maximum potential drop, due principally to the so-called constriction resistance.

To check the suitability of the previous current collector design, we have analyzed another design with the collecting tab in one corner [see Fig. 1 (d)]. Fig. 1 (d) shows a non- symmetric distribution of the equipotential lines and current streamlines. The potential gradient is also greater near the collecting tab, because of the convergence of streamlines and the increment of the in-plane current density. These results of our simulations are in agreement with those obtained in the literature[14-17]. This last geometry shows slightly higher maximum potential drop, nearly 14 % greater than that of the symmetric current collector. In the asymmetric current collector, the electron pathways are longer; and thereby this causes the increment of the $V_{max}$.

The most striking feature of the electrical behavior of these two geometries of current collectors is that they follow the same logarithmic trend as for the rectangular sheets [Eq. (4)]. Figures 2 (b) and (c) show the linear fits of the logarithmic relations for the symmetric and asymmetric current collectors ($R^2 = 1$). Therefore, the maximum potential drop of the current collectors is also inversely proportional to their thickness. Due to this relation, a current collector with a thickness of 10 μm shows a maximum potential drop 50 times greater than one of 500 μm (and 10 times greater than one of 100 μm). These increments of ohmic losses are important if we want to reduce the dimensions of any energy storage device.

Finally, we have checked that all the previous trends have no dependence on the material. As a result, we can verify that the critical thickness in which the maximum potential drop starts increasing has no dependence on the material. However, due to the greater resistivity of Al, the maximum potential drop for the same current collector of Al is approximately 60 % greater than in the Cu.

In conclusion, we have developed an electrical model to investigate the effect of the thickness on the electrical behavior of current collectors. For that purpose, current collectors of a commercial LIC were characterized. The maximum potential drop of current collectors shows the same ohmic relation with the thickness than a simple rectangular sheet. In addition, the electrical model shows that the centered collecting tab increases 50% the potential drop with respect a rectangular sheet. The displacement of the tab to a corner increases the maximum potential drop about 14 % with respect to the symmetrical configuration.

The information obtained in these simulations is extremely helpful for future prototyping, design and optimization of electrochemical energy storage cells (LIBs, supercapacitors, etc). We have proven that the thickness of current collectors is a critical design parameter. Reducing the thickness of current collectors, we can get different profits (greater density of energy or power and cheaper materials). However, as a result, the inner resistance of current collectors increases and dangerous heating could appear. Moreover, mechanical deformations of current collectors could happen too. It is clear that it is necessary to achieve an equilibrium between all the design parameters.

Authors are grateful to CIC Energigune for help in the inner characterization of the commercial LIC. Authors also wish to thank the support received from Mondragon Unibertsitatea, Mechanical and Industrial Manufacturing Department.

**Figure captions**

FIG. 1. Geometry, electric potential, equipotential lines and electric current streamlines for: (a) rectangular sheet with upward current crossing all the sheet, (b) rectangular sheet with current entering by the side faces and flowing upwards, (c) symmetric current collector and (d) asymmetric current collector. In all the four cases the material is Cu, $d = 10$ μm and $I = 21.17$ A.

FIG. 2. Maximum potential drop as a function of thickness for different intensities (Cu): (a) rectangular sheet with upward current crossing all the sheet, (b) symmetric current collector and (c) asymmetric current collector. Those values for thicknesses above 500 μm are nearly constant, and there have not been plotted.

FIG. 3. Logarithmic dependency of the maximum potential drop and thickness (Cu): (a) rectangular sheet with current crossing all the sheet, (b) symmetric current collector and (c) asymmetric current collector. Symbols: □ 2.94 A, ▲ 5.88 A, ○ 11.76 A and ♦ 21.17 A.

**Figure 1**

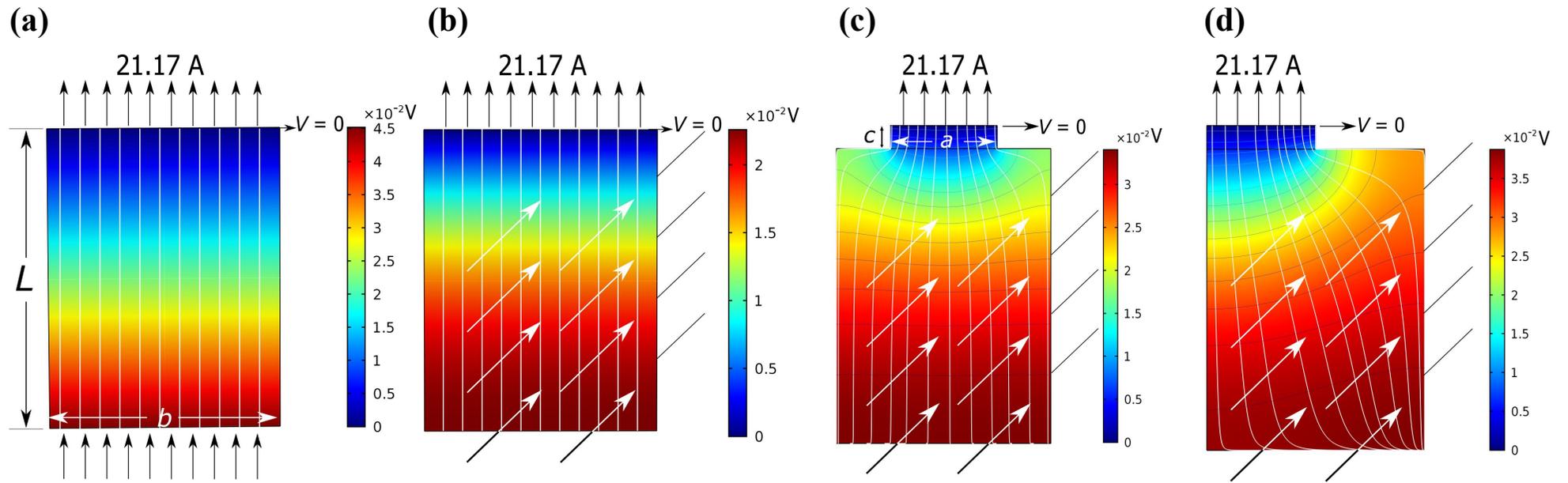

**Figure 2**

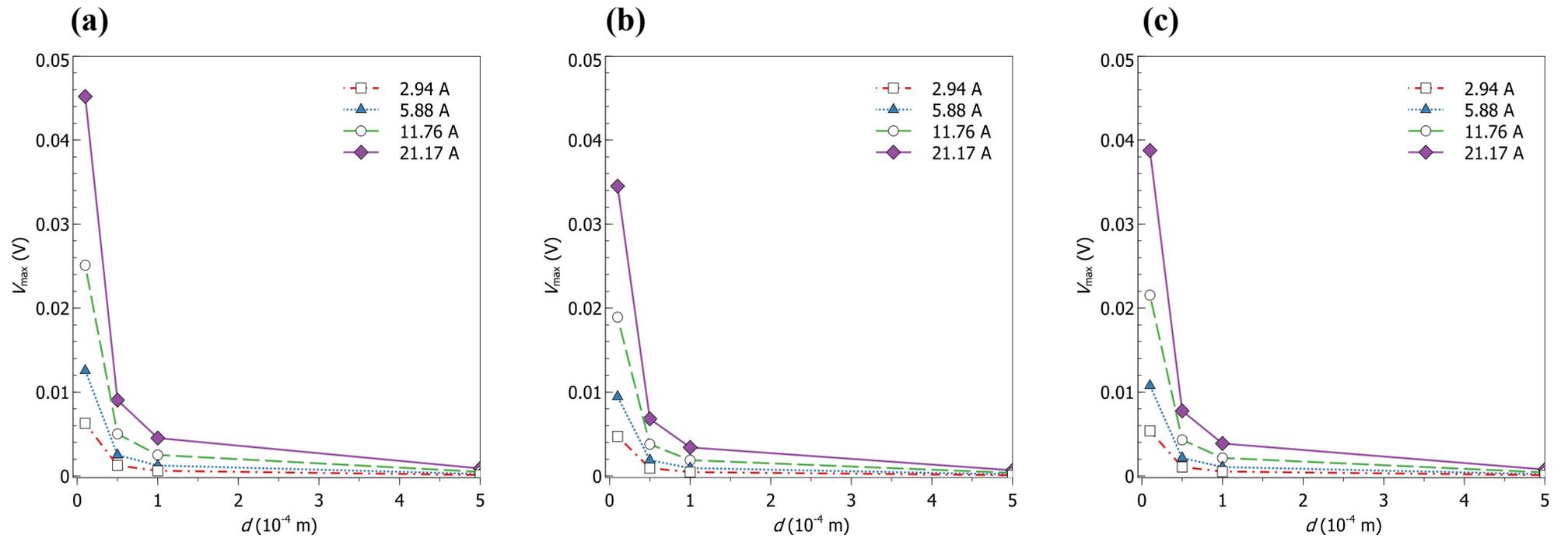

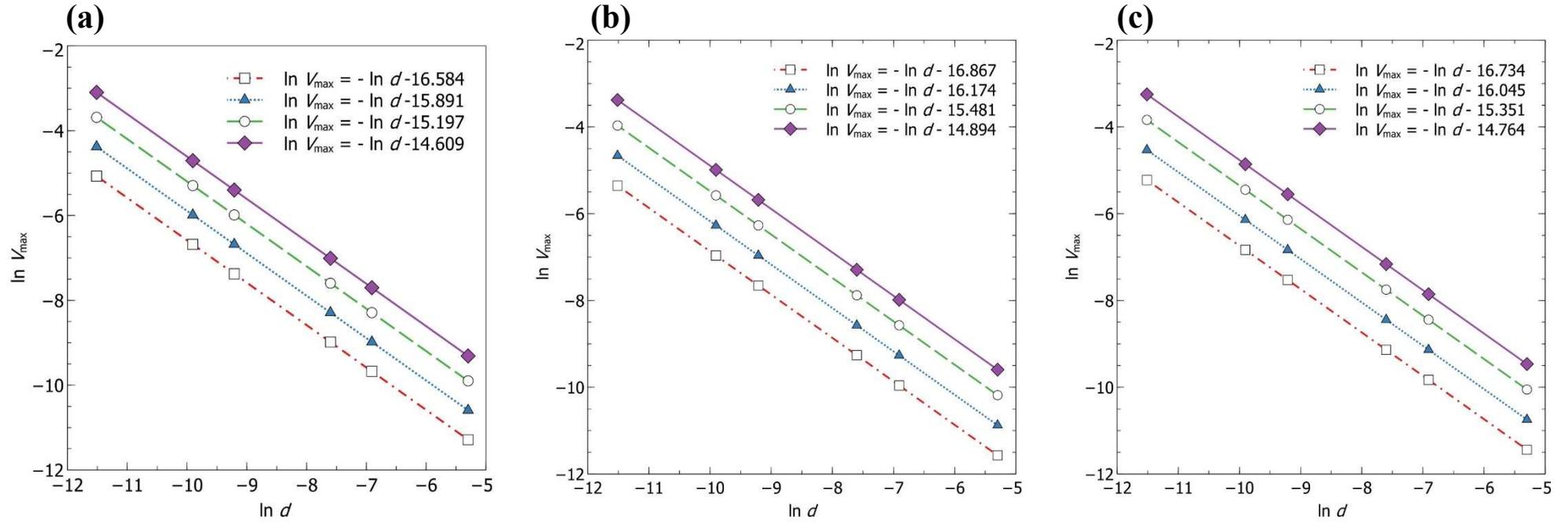